# SOA Governance – Road into Maturity

**George Joukhadar**
School of Information Systems, Technology and Management
University of New South Wales
Sydney, Australia
Email: g.joukhadar@student.unsw.edu.au

**Fethi Rabhi**
School of Computer Science and Engineering
University of New South Wales
Sydney, Australia
Email: f.rabhi@unsw.edu.au

## Abstract

There is a general consensus that SOA benefits could be reached but it is unclear how to achieve this. Research shows that the problems with SOA governance in practice are among the major reasons of SOA failures. Based on a literature review, this study first proposes a list of SOA aspects to be considered when implementing SOA governance. By adopting an interpretive research methodology based on interviews, this research paper makes two contributions: it addresses the practical matters that are major concerns for organisations to achieve a higher maturity level with their SOA, and it reveals the importance of the key SOA aspects in building strong governance and consequently reaching a higher maturity level. The expected result should deliver a theoretical contribution to SOA maturity in relation to SOA governance; it could provide organisations with new awareness in assessing their level of maturity and provide recommendations.

**Keywords** SOA Governance, IT governance, SOA Maturity, SOA Aspects, governance framework.

## 1 Introduction

It is apparent that industry readiness plays a major role in the decision to adopt Service Oriented Architecture (SOA) (Choi et al. 2010), and firms adopting SOA quickly can sustain or secure competitive advantage (Lee et al. 2010). According to *The Open Group*, SOA is an architectural style that supports service-orientation. Service-orientation is a way of thinking in terms of services and service-based development and the outcomes of services (TheOpenGroup 2015). Weir and Bell (2013) defined SOA as "strategy for constructing business-focused software systems from loosely coupled, interoperable building blocks (called services) that can be combined and re-used quickly, within and between enterprises, to meet business needs." (Weir and Bell 2013)

The implementation of SOA has nothing to do with technology platform that the organisation uses. It is widely recognized that the real difficulty lies in dealing with people and processes from different parts of business and aligning them to deliver enterprise wide solutions (Weir 2013). Even though many organisations approached SOA as simply implementing web services and some of them thought that SOA was a new integration technology that solve their IT complexity. They paid little attention to SOA governance and many organisations which built hundreds of services ran into different sorts of problems. According to Gil Long - IBM Distinguished Engineer and SOA governance integration lead, "SOA governance is about establishing a strategic vision for SOA, aligning that business and IT vision, sponsoring and funding that vision, providing responsible oversight and controls, and ensuring governance mechanisms are in place to support the vision." (Laningham 2007)

Research found that the implementation and formalisation of SOA governance is an essential phase for organisational maturity in SOA (Mahadevan et al. 2009; Weir 2013) and the evaluation of the level of maturity by implementing a SOA maturity model reflects the implications on organisational governance (Rathfelder and Groenda 2008). SOA maturity is a method of evaluating the organisation that creates an understanding of the maturity level of SOA within the organisations and its readiness to ensure that SOA governance framework is defined in an appropriate level for the organisation (Hojaji and Shirazi 2010c). Moreover, governance has been seen an essential phase in maturing the SOA of the organisation, because governance is about aligning people, processes and tools and creating an accountability framework to formalize SOA implementation by ensuring that everyone understands the contributions expected of them and the value they can add to the solution (Weir 2013).

The rest of the paper is organised as follows: Section 2 provides a background on SOA, SOA governance and SOA maturity, states the research gap and addresses the research question(s), section 3 presents the methodology used to address the research questions, section 4 addresses the findings while the last section presents the conclusion and future work.

## 2　Related Research

The implementation of SOA depends on the organisation itself (Woolf 2006). Although there are some essentials that organisations should be aware of: how SOA implementation will impact on the organisation and that managers should have an understanding of the business value and implementation benefits of SOA (Smith 2008).

One of the known challenges of SOA is that it requires people to work together in new and different ways; establishing a strong culture of governance could help overcoming this common challenge (Woolf 2006). Another type of the common challenges that organisations face is that service-oriented design and implementation are not governed properly or governed assuming the traditional software development without any attention paid to service orientation methodologies (Bernhardt and Seese 2009; Luthria 2009). We conclude that for organisations to address the implementation challenges they should start with governance (Hassanzadeh et al. 2011; Lundquist 2009; Niemann et al. 2009; Parachuri et al. 2008; Smith 2008; SoftwareAG 2008).

However, there are also some difficulties that organisations face in deploying SOA governance: difficulties in designing effective decision structures, managing and governing services, integration of legacy applications, lack of service funding, lack of consistent governance processes, building a SOA roadmap and assessing the SOA maturity (SoftwareAG 2008).

The adoption of a SOA governance framework could help overcoming the governance difficulties. The role of the framework is to focus on the architectural concepts, components and standards that are required to build effective SOA governance (Hojaji and Shirazi 2010c; Hojaji and Shirazi 2012).

There have been many proposed SOA governance frameworks; they all differ in design and on how to approach SOA initiatives. Most of them are based on 'theories' and assumptions rather than practice. There are no guidelines to adopt most of these frameworks and no empirical evidence regarding their implications and their success rate. There is so much that is not known and not researched and there are many claims made in the literature that are not substantiated by empirical evidence. As a result there is confusion about the usage of SOA governance, and hence the aspects of SOA have not been successfully addressed and not supported with empirical data. Reviewing the literature, we find conflicting claims regarding the coverage of these frameworks. What emerges as a critical issue in any adoption of SOA governance is not so much which governance framework to choose, but more importantly to identify and focus on SOA particular aspects that need to be addressed irrespective of the framework.

There is also a general consensus that there are many different aspects of SOA and they vary across different authors. The aspects are considered as "the most essential topics in SOA governance to be placed in a framework to ensure a consistent transition from strategic considerations to delivery of the services" (Hojaji and Shirazi 2010c). This research paper therefore first examines a range of aspects of SOA that are of key importance when an organisation adopts an IT or SOA governance framework. Based on an extensive literature review on SOA and IT governance frameworks, this research initially proposed a list of key SOA aspects based on previous researches (Boerner and Goeken 2009; de Leusse et al. 2009; Derler and Weinreich 2007; Hojaji and Shirazi 2010a; Kuang-Yu et al. 2008; Larrivee 2007; Niemann et al. 2008; Varadan et al. 2008). The aspects were presented in a previous work for the same authors (Joukhadar and Rabhi 2014). This list of aspects adopted in this research considered the following: SOA strategic vision, SOA roadmap, SOA maturity, service lifecycle management, service portfolio management, SOA business capabilities, governance processes, organisational change management, SOA governance board, Centre of Excellence (CoE), open boundaries management, service performance analysis, policy management, best practices deployment, SOA governance technology, infrastructure capability, process monitoring and evaluation, service transparency control, and service security control. The importance of SOA maturity and its association with SOA governance have been addressed in the literature and are the core of this research paper.

According to TheOpenGroup, the purpose of the assessment of SOA maturity is to create an understanding of the maturity level of SOA within the organisation and its change readiness to ensure the SOA governance framework is defined to a level appropriate for the organisation. Most importantly when the maturity level increases, the SOA governance regimen need to be revised and modified accordingly. The SOA governance roadmap is formed by first assessing the maturity level of the organisation, and then putting complementary effort to define where the organisation wants to be. The Open Group had a SOA maturity model called Open Group Service Integration Maturity Model (OSIMM) to structure the assessment of SOA (TheOpenGroup 2009).

Niemann et al. (2008) included SOA Maturity Measurement (SMM) in their framework and argued that its role is to assess the readiness and maturity of the SOA processes, providing feedback to the organisational entities in the organisation. "An SMM assesses an Enterprise Architecture (EA) system in terms of its SOA conformance. It defines several maturity levels, where each level defines goals and metrics which determine and verify the current maturity level of an SOA implementation" (Niemann et al. 2008); Niemann et al. (2009) added that the achievement of governance goals is supported by SOA Maturity Models and respective governance mechanisms (Niemann et al. 2009).

Hojaji and Shirazi (2010b) argued that when adopting SOA, the governance framework enables organisations to manage complexity, to improve the ability to make better decisions, and to develop the necessary maturity and infrastructure in the form of control and enforcement mechanisms (Hojaji and Shirazi 2010b); therefore the governance framework should provide metrics and maturity models to assess the level of SOA maturity (Hojaji and Shirazi 2010b). For organisation to assess where they are in the migration path to SOA, Hojaji and Shirazi (2010b) proposed two maturity models in their SOA governance frameworks in order for organisations to achieve SOA benefits with higher levels of maturity: process maturity model and SOA adoption maturity model (Hojaji and Shirazi 2010b).

To the best of our knowledge, no publication has provided empirical guidelines that could be helpful for organisations on how to deploy the governance framework or the maturity model. There is a lack of empirical studies around the issues of adopting SOA governance. In particular there is no clear understanding about the adoption of a SOA governance framework and how it can help organisations achieve a higher level of SOA maturity. This research paper focuses on SOA maturity as a crucial aspect of SOA adoption to be integrated in the SOA governance framework. It addresses the concerns that organisations face with current governance frameworks and provides recommendations for organisations to build up their SOA governance frameworks and then achieve the desired maturity level. The research questions that this paper addresses are:

- What are the pragmatic concerns that organisations face when implementing SOA?
- How can organisations implement a SOA governance framework and which SOA aspects of the framework help achieving a higher level of maturity?

## 3    Methodology Used and Data Analysis

Since the research problem statement is concerned with determining a common understanding of what SOA governance and SOA maturity means in practice by highlighting the aspects of SOA governance frameworks, this research uses an interpretive qualitative research methodology based on interviews as a first phase to address the research questions.

Qualitative research approaches and methodologies are well established in IS research (Goldkuhl 2012; Walsham 2006). Walsham (2006) argues that interpretive research has clearly become much more important in the Information System (IS) field than it was in the early 1990s. He added "many IS journals are now publishing interpretive studies". Walsham and Goldkuhl both agreed that empirical data generation is seen as a process of socially constructed meanings that is socially constructed by researchers and participants. Following a long tradition in qualitative, interpretive research in IS, Goldkuhl (2012) claims that scientific knowledge should be based on the meanings and knowledge of the studied actors and also co-constructed through inter-subjective meaning by the actors and researchers making during the empirical study. Given that the focus of this study is the practice of SOA governance frameworks and finding how SOA governance can help implementing and maturing SOA in real-life contexts so as to create an understanding of the discrepancy between theory (how frameworks should be applied) and the practice, an interpretive methodology is adopted.

The first phase of the research involves conducting interviews with experienced industry professionals who have experience with SOA governance in multiple organisations and who participated in several SOA projects. Twenty eight interviews were conducted and the participants were selected and recruited through professional networks of SOA/IT governance professionals. The professional social network LinkedIn was used to contact candidates worldwide. Contacts were made by email and one page summary of the research was sent upon request. The first stage of the selection criteria was based on email communication to confirm that the candidates meet the selection criteria. For the research sake, the first part of the interview includes a set of questions asking about the participants background, experience in IT/SOA governance and service management.

The interviews took the form of face-to-face, Skype or phone interviews. Each interview has taken from 1 hour up to 2 hours. Most of the participants had a decision-making role in their organisations and their experience with IT/SOA governance varies from 4 to more than 25 years. They have worked with a minimum of two organisations and on different SOA projects in different sectors. Table 1 shows a snapshot of the participants' demographics. The first three interviewees were not recorded. They were pilot interviews that helped to adjust and evaluate the interview questions in order to capture additional empirical data and to try to achieve the aims of this research. They also helped in timing the interviews. Participants 19 and 20 decided to participate by email. All participants agreed to be audio-taped except participant 4.

|  | Job Title | Sector | Location | Years of exp. | Gender | Interview |
|---|---|---|---|---|---|---|
| P1 | Project Manager | Computer Software | Australia | 7 | M | Face-to-face pilot interview |
| P2 | Head of Enterprise Architecture | Government | Australia | >25 | M | Face-to-face pilot interview |
| P3 | Integration Architect/Solution Architect | Software Development / API Management | Australia | 4 | M | Face-to-face pilot interview |
| P4 | Enterprise Integration Architecture | Financial Services | Australia | 14 | M | Face-to-face |
| P5 | Solution Architect | IT Solution Services | Australia | >25 | M | Face-to-face |
| P6 | Governance Director | Education – Technology Management. | Australia | 15 | M | Face-to-face |
| P7 | Managing Director – SOA Governance consultant | Communication – API Management | South Africa | 15 | M | Skype |
| P8 | Chief Enterprise Architect | Financial Services | Australia | 15 | M | Face-to-face |
| P9 | Senior Solution Architect | Financial Services | Australia | 13 | M | Face-to-face |
| P10 | Information Technology Professional | Financial Services | Australia | >25 | M | Face-to-face |
| P11 | Architect / Consultant | Information Technology and Services | US / Canada | >25 | F | Skype |
| P12 | Technology & Solution Architect | Information Technology and Services | Australia | >25 | M | Face-to-face |
| P13 | Principal Architect | Information Technology and Services | Australia / US | 25 | M | Face-to-face |
| P14 | SOA Architect / Enterprise Architect | Software Development | Australia / US | 18 | M | Face-to-face |
| P15 | SOA Consultant / SOA Architect | Information Technology and Service | India | 5 | M | Skype |
| P16 | Enterprise SOA Architect | Information Technology and Services | US | >25 | M | Skype |
| P17 | SOA Governance specialist | Information Technology and Services | US | >25 | M | Skype |
| P18 | System Analyst | Retail Services | MEA | 4 | M | Skype |
| P19 | Executive Architect | Financial Services | Australia | >25 | M | Phone |
| P20 | Senior Solution Architect | Information Technology and Services | Europe | 5 | M | Email |
| P21 | Enterprise Architect | Information Technology and Services | India/MEA | >25 | M | Email |
| P22 | Enterprise SOA Architect | Financial Services | US | 17 | F | Face-to-face |
| P23 | System Analyst | Distribution | MEA | 10 | M | Skype |
| P24 | SOA Technical Architect | Computer Software | Canada | 20 | M | Skype |
| P25 | Software Development Manager | Financial Services | Australia | >25 | M | Skype |
| P26 | Principal Architect | Information Technology and Services | UK | 10 | M | Skype |
| P27 | Chief Technology Officer | Computer Software | US | >25 | M | Skype |
| P28 | Principal Consultant | Information Technology and Services | Australia | 10 | M | Face-to-face |

*Table 1 - Participants' demographic*

A broad set of questions were asked regarding the participants' background, their experience with IT and SOA governance, the role of each of the SOA aspects when implementing a SOA governance framework, the importance and role of each aspect in building strong governance. The strategy to be followed when implementing the governance framework took a part of the discussion.

The interviews' transcripts were analysed using Thematic Analysis - which is one of the more widely accepted methods of analysing qualitative data. The analysis was done by first coding interesting ideas, topics, and concepts and then organizing them into themes and identifying links among them. Coding in Thematic Analysis helps the researcher to build a systematic account of what has been observed and recorded (Ezzy 2002). The data analysis occurred over two stages: manually in the first stage and then using the qualitative data analysis software NVivo in the second stage. The manual data analysis helped in understanding the data codes and concepts and identifying new concepts. The research is still under development and at the finale stage of the data analysis.

## 4  Findings

The interviews discussed the role of SOA governance solution in helping organisations mature their SOA practices and thus ensuring SOA success. The interviewees expressed their views and experience with the aspects of SOA that are required for a governance framework. They were asked to discuss each of the aspects mentioned in section 2, the interconnection between these aspects, their impact on each other and the role of each aspect in building a governance framework. The participants' views to SOA were based on their own experience. Most of the participants agreed that SOA needs a customised governance framework. Therefore, they either modified the selected framework to match with their organisational needs or built their own one. For each aspect, participants were asked to assess its importance and its practical usage by giving it a score between 1 (lowest) and 5 (highest). It was argued by the majority of participants that the business should drive the need for maturity, however it was noted that it is often not done effectively by many organisations for different reasons. This section presents the views and findings about the practical challenges that organisations face and preventing them from achieving a higher level of maturity. Then it provides recommendations showing how implementing SOA governance can help organisations customise the SOA aspects of the framework and reach the required maturity level.

### 4.1  SOA Maturity Concerns

Some participants consider SOA maturity assessment less important at the beginning of the SOA implementation and therefore gave it a score of 4 or 3 instead of a 5. Those participants noted that SOA maturity assessment becomes more important as time goes on and then will get a score of 5. SOA Maturity was given an average of 4.1 for its importance and 2.95 for its practical usage. According to the participants of this research the gap between the importance and practical usage scores is because not a lot of companies are maturing their SOA governance and they misunderstand the concept of SOA: Participants 4 and 6 stated that some organisations think that they are doing SOA, but they end up doing systems integration or enterprise application integration instead, while participants 7 and 15 argued that some organisations are still confused about what the services are, how they should be managed and how they fits into the bigger picture. Participant 7 added that some organisations don't realize that services need to be managed and that they are fictional blocks for an organisation which they can use.

The second concern that was mentioned by participants 6 and 8 is project management. In many situations SOA governance becomes more like a project management or is affected by project management. 'Project-focus', 'short term projects' and 'quick wins' are some the most challenging facts that affect the implementation of SOA.

Overestimating the SOA maturity level is another concern because some organisations tend to think that they are quite mature, and they approach SOA as a big bang approach according to participants 7, 14, 15 and 26. Participants 6 and 7 argued that organisations should be aware that SOA governance is still evolving and not mature enough when compared to other domains where Enterprise Architecture (EA) and Information Technology (IT) governance are looking at.

The concerns with existing frameworks were confirmed by participants 4, 7, 8, 15 and 22; the existing SOA governance frameworks are very theoretical and they don't have a pragmatic implementation. It was recommended by the majority of participants that frameworks should be custom-tailored or used as a reference only. According to participants 6, 7 and 15 there are always elements in the framework that are not applicable for an organisation; therefore organisations shouldn't be rigid and try to follow the framework because of the framework. Regarding vendor-based and open-source frameworks, participants 3, 4, 6-18, 20-28 agreed that vendors dictate to their customers that they need to go with their own products. Simultaneously, some organisations have got inclination towards some open source tools for implementing governance framework. Open source could be problematic for large organisations according to participant 8. Moreover, current SOA governance frameworks do not support Application Programming Interface (API) portals yet as mentioned by participants 7, 13, 15 and 17, and they do not support flexible and virtual Enterprise Service Bus (ESB). The need for a virtual ESB is claimed by participants 8 and 15 that with the presence of the cloud, organisations might need part of their services in the secure cloud, a second part on site, and another part shared with a partner or a supplier or both.

Aligning SOA governance framework with IT governance and Enterprise Architecture (EA) governance could be problematic to some organisations as it depends on the maturity level of the organisation and on the governance body and governance structuring of the organisation according to participants 11, 15 and 25. Participants 4, 6, 15 and 26 claimed that many organisations do not see SOA governance as an extension to EA governance in addition to being an extension to IT governance.

According to participants 7, 13 and 15, some organisations that understand SOA have problem reusing services especially when moving to cloud computing. This is because they do SOA without having policy management and they do not implement run-time governance properly; they do it at the expense of design-time governance, therefore they struggle to achieve a higher level of maturity.

### 4.2 Achieve a Higher Maturity Level

Most participants agreed that applying SOA governance appropriately is the key for organisations to understand SOA and achieve a higher SOA maturity level. It was argued that SOA governance should also be part of the Enterprise Architecture (EA) governance. This will help enforcing the framework implementation according to participants 7, 9 and 26.

According to participants 5, 7, 13 and 26, organisations are not mature at the start; nevertheless organisations should have a clear vision of where they want to be and understand what the approach needs to be. Participants agreed that organisations should have a very mature governance/architecture board as part of their enterprise; that board should build a well-defined vision and roadmap to succeed in their SOA journey. The maturity level is first assessed at the beginning of the SOA journey: what is the current state of SOA maturity, i.e. where the organisation is standing in its SOA maturity, and what the organisation is seeking to achieve. These two activities serve as input to build the formation of the SOA roadmap; the roadmap role is to take the organisation from the current position to the new expected one with the presence of SOA vision. Participants agreed that organisations have to understand and know where they need to go, and should have an understanding of their maturity. According to participants 5, 10, 11, 13, 15, 17 19, 24, 26 and 28 organisations need to look at two or three weeks of assessment exercise workshop done by SOA professionals to assess the maturity level and recommend to them the maturity roadmap. The assessment is done by looking at the maturity level in terms of service integration, the current infrastructure and the governance processes they have, the methodologies they follow, the kind of tooling they are using, etc. SOA maturity gets more important as the implementation goes on. When the maturity level increases the SOA governance framework need to be reviewed and modified; hence organisations should expect their SOA maturity to go up as time goes on.

Participants also agreed that organisations should rely on the years of experience they have to build their SOA maturity roadmap against the framework they deploy. This could be done by applying best practices, change management solutions and deploying a Centre of Excellence (CoE).

Best practices deployment depends on the understanding and repeatability of the organisation. Organisations become more mature when they are able to understand why they are doing SOA, why they need to do it and how they are doing it. They should follow pragmatic solutions and learn from failures. According to participants 4, 13, 24, 25 and 26, the only way for organisations to become mature is to have some failures, or learn from people that have had failures to import that maturity. "A lot of time maturity comes after two or three times of working with things" claimed participant 13. Depending on their experience, some organisations take a whole life cycle approach and have a fairly strong process to do SOA governance, not just at an investment appraisal level. That sort of maturity takes sometimes around four or five years into the SOA governance process according to participant 6. Most participants argued that the framework is only a reference; the roadmap is the implementation, how organisations should use that framework, and how they should go up in their SOA maturity. For example, participant 7 uses what they call "enablers" – an enabler is defined as the solution to solve a specific pain point or problem for an organisation - that help to get to the next maturity level. Participants 6, 7, 10, 13 and 22 advised to stay away from written documents that get onto the shelf and never been used, but organisations should rely on their expertise.

Another important factor to achieve higher maturity according to participants 6, 13, 17, 25 and 26 is to have a proper organisational change management to prepare the right culture. Organisations will never reach maturity if they don't have the right culture to accept the changes required when implementing SOA; this is because SOA change how the organisation work "maturity needs to go hand in hand with culture" according to participant 26.

The need for a Centre of Excellence (CoE) depends on how large the organisation is. The CoE would only ever come when organisations are mature enough to understand what it is for. According to participants 5, 7, 8, 9, 11-15, 25-26, the larger the organisation is, the more mature it becomes in SOA governance and the more need for a CoE to enforce best practices, guidelines, etc. According to participant 9, the CoE is needed when the SOA maturity for an organisation is low and where a capability is required to be built inside that organisation. The goal is that everyone in the organisation have a role in SOA and do SOA as part of their daily work. When setting up the CoE, participant 7 argued that organisations need to setup service lifecycles and they can see improvement of the services portfolio and services documentation, and the people in the organisation start using the governance processes.

Based on the views of participants 7, 11, 13 and 15 and 24-26 organisations should consider run-time governance separately from design-time governance to overcome the problem of reusing services. At design-time and development-time governance, SOA governance technology is considered less important compared with run time governance and tools are required to get up with high maturity. The run-time governance depends highly on the changing environment inside the organisations; business change involves technology change and in that case it depends on how organisations apply policy management. Participant 14 argued that service governance also helps organisations to drive the maturity; service governance should have its own function because it requires a different set of skills and it almost needs to be fine-tuned per service.

## 5   Conclusion and Future Works

We conclude that SOA governance only works when it reflects what organisations are doing and what they have agreed to do. The agreed cultural change, the agreed process and the agreed behaviour that organisations develop are reflected in governance. Therefore governance can only work if it reflects that reality. The aspects of SOA governance listed in the second section of this paper were examined by the participants of this research. The key focus was to determine the role of the governance framework and the role of each of SOA aspects in building effective SOA governance during SOA adoption.

In section 4 we mentioned most of the SOA aspects listed in section 2: SOA vision, SOA roadmap, SOA maturity, governance board, Centre of Excellence, service lifecycle management, governance processes, organisational change management policy management, best practices, governance technology and others that the SOA governance framework is composed of. This paper addressed the issues that organisations face when trying to go higher in their SOA maturity and presented suggestions for organisations to consider; it presented the role and importance of those aspects in establishing strong SOA governance and therefore achieving a higher maturity level. Participants of this research agreed that the role of SOA maturity is to drive an improvement on how organisations do things where governance should always reflect how those things are done, and though SOA maturity assessment has been considered very important for SOA governance. Therefore the adoption of a governance framework depends on the maturity level of the organisation in terms of SOA infrastructure and whatever processes they have been following. Accordingly and depending on the current state of the maturity level, organisations should consider the target maturity level in terms of governance state. The findings cited in this paper could help organisations consider which aspects they need to deploy in a SOA governance framework from the beginning of the SOA implementation to achieve a higher maturity level and not after the technology is applied.

Phase I of this research is at the final stage of data analysis and promising more results. When Phase I is completed, two organisations will be selected for Phase II for an in-depth field research study: one organisation that has been highly successful with SOA governance and another one that is starting to implement SOA governance. Comparing and contrasting the results of the two phases will provide grounding for the development of substantive theoretical claims regarding the importance and role of SOA maturity assessment as an essential aspect in building an effective SOA governance framework. On the theoretical level, this study is expected to contribute to conceptualizing the aspects of SOA and the interconnections between them. On the practical level, the results should help in assessing the practical deployment of SOA governance frameworks and their effectiveness; it will help SOA Governance Boards to evaluate and update their governance frameworks. It will also provide guidelines for organisations when they consider adopting a SOA governance framework.

# 6　References